\documentclass[twocolumn, twocolappendix]{aastex631}

\usepackage{amsmath}
\usepackage[T1]{fontenc}

\newcommand{\BIBORDERINGTAG}[1]{}

\shorttitle{Dust Torus Spectra from Multi-Scale AGN Simulation}

\begin{document}

\title{Early Stages of Dusty Tori: The First Infrared Spectra from a Highly Multiscale Quasar Simulation}

\author[0009-0002-8417-4480]{Jaeden Bardati}
\affiliation{TAPIR, Mailcode 350-17, California Institute of Technology, Pasadena, CA 91125, USA}
\email{jbardati@caltech.edu}

\author[0000-0003-3729-1684]{Philip F. Hopkins}
\affiliation{TAPIR, Mailcode 350-17, California Institute of Technology, Pasadena, CA 91125, USA}

\author[0000-0002-1061-1804]{Gordon T. Richards}
\affiliation{Department of Physics, Drexel University, 32 S. 32nd Street, Philadelphia, PA 19104, USA}

\begin{abstract}
We present the first infrared spectral predictions from a self-consistent simulation of the formation of a quasar in a starburst galaxy, spanning the cosmological environment to scales well below the dust sublimation region. The infrared (IR) emission is dominated by a torus-like dust structure composed of the highly magnetized, turbulence-supported outer accretion disk and of accreting gas tidally torn from the interstellar medium (ISM). At these early stages, the active galactic nuclei (AGN) is buried and Compton-thick. The near- to mid-IR escaping luminosity varies by almost an order of magnitude across sightlines, largely due to extinction from the inflowing stream of cold dust. Self-absorption within the torus suppresses silicate emission features, and further reprocessing by the ambient ISM leads to prominent silicate absorption and colder IR emission. The sublimation structure is stratified by composition and size, producing sightline-dependent extinction curves that intrinsically vary in shape. However, after repeated scattering in the optically thick dusty medium, these curves emerge substantially grayed. We also demonstrate that bipolar outflows from the central black hole that carve biconical cavities and reveal the central engine in later stages can preserve IR anisotropy and silicate features. These results suggest that dusty starburst quasars can undergo a buried, IR-bright phase early in their evolution.
\end{abstract}

\section{Introduction} \label{sec:intro}

Supermassive black holes (SMBHs) are thought to exist in the centers of most massive galaxies \citep{Begelman_1980, Volonteri_2003, Burke_2025} and have masses that correlate with many properties of the bulge in its host galaxy such as mass \citep{Magorrian_1998}, velocity dispersion \citep{Ferrarese_2000, Gebhardt_2000}, morphology \citep{Graham_2001}, and binding energy \citep[]{Aller_2007}, among others \citep[see the review by][]{Kormendy_2013}. These relations suggest that SMBHs and their host galaxies co-evolve by active galactic nuclei (AGN) feedback mechanisms \citep{Silk_1998, King_2003, DiMatteo_2005, Hopkins_2005, Torrey_2020}, thought to manifest via radiation, outflows, or relativistic jets from the inner accretion disk \citep{Crenshaw_2000, Fabian_2009, Dunn_2010, Sturm_2011, Tchekhovskoy_2011, Zakamska_2016, Williams_2017, Veilleux_2023, Vayner_2024, Liu_2024}. This feedback occurs during a few short phases of active accretion \citep{Soltan_1982, Salucci_1999, Yu_2002, Wyithe_2003, Hopkins_2006}, when the structure of the AGN is thought to roughly follow the standard AGN unification model \citep{Antonucci_1993, Urry_1995}, explaining most of the AGN diversity using only the accretion rate and relative orientation. Crucial to the unification model is the dust torus, which obscures the accretion disk and broad line region (BLR) emission along a covering fraction of sightlines and re-radiates in the infrared \citep{Sanders_1989}. In this model, type 2 AGN are explained as viewing the central engine through lines of sight obscured by the dust torus, whereas type 1 are the unobscured sightlines. The dust "torus" is unresolved in all but a few local galaxies \citep[e.g.][]{Tristram_2007, Raban_2009, Tristram_2014, GRAVITY_2020, GRAVITY_2024}, instead requiring detailed radiative and hydrodynamical modeling to constrain its underlying physical conditions. 

Although the dust torus is central to AGN unification, its physical origin, structure and dynamics remain poorly constrained.
Since the dust torus lies near the boundary between the regions whose structure and dynamics rely on physics related distinctly to the central engine (e.g. BLR) or those related to the interstellar medium (ISM, e.g. narrow line region), it is not clear which regime dominates. The obscuring region could compose of dust in the outer accretion disk, where the inner radius is set by the sublimation temperature of the dust \citep{Barvainis_1987, Pier_1992, Kishimoto_2007}, or it could lie in a region beyond the circularized accretion disk, at sufficiently large radii where self-gravity dominates (Toomre $Q \lesssim 1$) leading to a torus composed of fragmented dust clouds orbiting the central SMBH \citep{Wada_2002, Thompson_2005, Wada_2009, Hopkins_2012}. 
There is also the question of what physical mechanism supports the dust structure from collapsing to a scale height $H/R \ll 1$ expected of standard thermal-pressure-dominated \citet{Shakura_1973} disks, but is inconsistent with the $H/R \gtrsim 1$ expected from measurements of the covering factor \citep{Lawrence_1991, Maiolino_2007, Lusso_2013, Stalevski_2016}. Various vertical support mechanisms have been proposed including stellar feedback \citep{Fabian_1998, Wada_2002, Zier_2002, Thompson_2005, Nayakshin_2007, Schartmann_2009, Hopkins_2012} or strong magnetic field support \citep{Begelman_2007, Gaburov_2012, Forgan_2017, Mishra_2020, Mishra_2022, Liska_2024, Mishra_2024, Hopkins_2024d}, but it is still not obvious which processes dominate and under what circumstances. 
It is also not clear what the morphology of the torus is, such as whether it is smooth \citep{Granato_1994, Schartmann_2005, Fritz_2006}, clumpy \citep{Rowan_Robinson_1995, Nenkova_2002, Honig_2006, Dullemond_2005, Elitzur_2006, Nenkova_2008, Schartmann_2008}, or some "two-phase" combination \citep{Stalevski_2012, Siebenmorgen_2015, Stalevski_2016}.
Finally, the dynamical state of the dust torus is equivalently poorly constrained and possible dynamical scenarios include the dust torus containing or composing of an outflowing wind \citep{Honig_2017, Stalevski_2017, Lyu_2018, Stalevski_2019, Venanzi_2020}; inflowing material \citep{Wada_2002, Thompson_2005, Hopkins_2012}; material purely rotating or turbulently held in place \citep{Wada_2012, Schartmann_2014, Wada_2016}; or some mixture of each. Recent IR interferometry and Atacama Large Millimeter Array (ALMA) observations on local Seyfert galaxies tend to support such a mixed picture, revealing both compact rotating dust and molecular gas structures and extended polar emission likely associated with outflows \citep{Honig_2012, Lopez_Gonzaga_2016, Asmus_2016, Garcia_Burillo_2021, Alonso_Herrero_2021, Garcia_Bernete_2022}.

Despite substantial progress in exploring and constraining the structure and dynamics of the dust torus, the problem is highly multi-scale and multi-physics dependent, making it difficult to model the full system self-consistently and with realistic initial conditions. Recently, \citet[][henceforth FORGE'd in FIRE]{Hopkins_2024a} simulated the formation of a quasar accretion disk starting from cosmological initial conditions. The simulation zooms in on a massive galaxy during merger and starburst, at a redshift of $z\sim 4.4$, tracking the material inflow down to the central $1.3 \times 10^7 M_\odot$ black hole to a distance of $\sim 10^{-4}$ pc, including resolving individual star formation and all relevant physics self-consistently. Super-Eddington accretion rates of $\dot{M} \sim 10 - 100~M_\odot/\mathrm{yr}$ are sustained down into the accretion disk, driven at large scales by gravitational torques between gas and stars until star formation is dramatically shut down at $\lesssim 1~\mathrm{pc}$ from the SMBH by increasing optical depths to cooling radiation and strong magnetic fields. In the transition region, there is a magnetic pressure-dominated gravitoturbulent accretion disk, distinct from traditional $\alpha$ disks or magnetically arrested disks \citep{Hopkins_2024d}. The strong torodial magnetic fields supporting the outer accretion disk are naturally produced by advection from the surrounding ISM cloud complex \citep{Hopkins_2024b}. However, a detailed radiative analysis is required to understand the nature and dynamics of the obscuring dust structure in this system. 

In this paper, we present the first infrared spectral predictions from a fully self-consistent active galactic nucleus simulation starting with cosmological initial conditions and the first infrared spectra from a magnetically-dominated disk system. Specifically, we study the continuum emission from running the post-processing radiative code SKIRT on FORGE'd in FIRE when including a fiducial central accretion disk source.
In Section \ref{sec:sim} we describe our simulations used, including a brief summary of the FORGE'd in FIRE simulation (\S \ref{subsec:fif}), as well as our post-processing radiative transfer simulation with SKIRT (\S \ref{subsec:skirt}). In Section \ref{sec:results}, we describe and discuss our results, including describing the basic torus structure and support mechanism (\S \ref{subsec:whatistorus}), the emergent spectra along various sightlines and scales (\S \ref{subsec:spectra}), the origin of anisotropy persistent through to the MIR (\S \ref{subsec:anisotropy}), the luminosity-dependent sublimation structure stratified by dust size and composition (\S \ref{subsec:substructure}), the diverse extinction curves grayed by the surrounding optically thick medium (\S \ref{subsec:extcurves}), 
the effect of re-processing by the ambient dusty ISM (\S \ref{subsec:extendingout}) and of evacuating a bicone as a preliminary analysis of to the impending polar outflow suggested by other studies (\S \ref{subsec:outflow_analysis}), as well as some general caveats (\S \ref{subsec:caveats}). 
We summarize our conclusions in Section \ref{sec:conclusion}. 

\section{Simulations} \label{sec:sim}

\subsection{Initial FORGE'd in FIRE Simulation} \label{subsec:fif}

We use the FORGE'd in FIRE simulation \citep{Hopkins_2024a}, a radiation-magnetohydrodynamics zoom-in from cosmological initial conditions to a $M_\mathrm{BH} = 1.3 \times 10^7 M_\odot$ quasar, based on the FIRE-3 \citep{Hopkins_2023FIRE3} and STARFORGE \citep{Grudic_2021} physics in the GIZMO meshless finite-mass (MFM) code \citep{Hopkins_2015}. The simulation begins with a 100 cMpc per-side cartesian coordinate box at $z \sim 100$ and evolves to $z \sim 4.4$ where a galaxy merger induces a large inflow of material to the center of the starburst galaxy. The simulation is then refined again and again until it reaches $\sim 300$ Schwarzschild radii ($\sim 10^{-4}$ pc) from the central black hole, resolving into the galactic core, through the end of star formation, past the dust sublimation region and into the outer regions of the accretion disk. The simulation contains the most important physics at these scales, including full self-gravity with adaptive softening and high-order Hermite integrators to accurately track many orbits \citep{Grudic_2020, Grudic_2021b, Hopkins_2023}; magnetohydrodynamics with kinetic effects from anisotropic Spitzer-Braginskii conduction and viscosity \citep{Su_2017, Hopkins_2017}; multi-band radiation hydrodynamics with adaptive-wavelength bands \citep{Hopkins_2019, Hopkins_2020}; thermo-chemistry of dust, molecular, atomic, metal-line, and ionized species coupled to the radiation and including metal enrichment and dust destruction \citep{Colbrook_2017, Grudic_2021, Choban_2022}; SMBH seeding and accretion \citep{Hopkins_2016, Shi_2023, Wellons_2023}; and individual protostar and star main-sequence evolution, mass loss, supernovae and jets coupled to the thermo-chemistry and radiation \citep{Grudic_2021, Grudic_2022, Hopkins_2023FIRE3}. We note that, in this paper, we do not use the recent version of FORGE'd in FIRE that continues the zoom down to the innermost stable circular orbit (ISCO) \citep[i.e.][]{Hopkins_2025} since we are primarily interested in the dust, which is entirely sublimated in that extra level of refinement. We only briefly summarize the most salient details of FORGE'd in FIRE here and refer the reader to \citet{Hopkins_2024a} for more details.

Particularly relevant to this paper is the implementation of thermochemistry, dust and radiation in FORGE'd in FIRE. In particular, all the dominant chemical processes at temperatures of $T \sim 1 - 10^{10}~\mathrm{K}$ are included, using non-LTE ion/atomic/molecular chemistry, heating and cooling. FORGE'd in FIRE also includes non-equilibrium dust that is scaled according to a $f_\mathrm{dtg} = 0.01 (Z/Z_\odot) \exp(-T_\mathrm{dust}/1500~\mathrm{K})$ dust-to-gas ratio, allowing dust destruction around $T_\mathrm{dust} \gtrsim1500~\mathrm{K}$. These processes are directly coupled to the explicit radiation treatment in the simulation. The radiation in FORGE'd in FIRE is handled through M1 radiation-hydrodynamics \citep{Levermore_1984, Hopkins_2019, Hopkins_2020} in five wavelength bands (extreme UV, FUV, NUV, optical/NIR, and an adaptive IR blackbody). Appropriate opacities for all included species with number densities $n \sim 10^{-10} - 10^{15}~\mathrm{cm}^{-3}$ are included, such as ion/atomic/molecular/metal lines, bound-bound, bound-free and free-free emission, as well as Compton and Rayleigh scattering. The gas and dust opacities are separately accounted for in the different radiation bands according to the gas, dust and radiation temperatures.

\begin{figure*}[t]
    \centering
    \includegraphics[width=\textwidth]{ 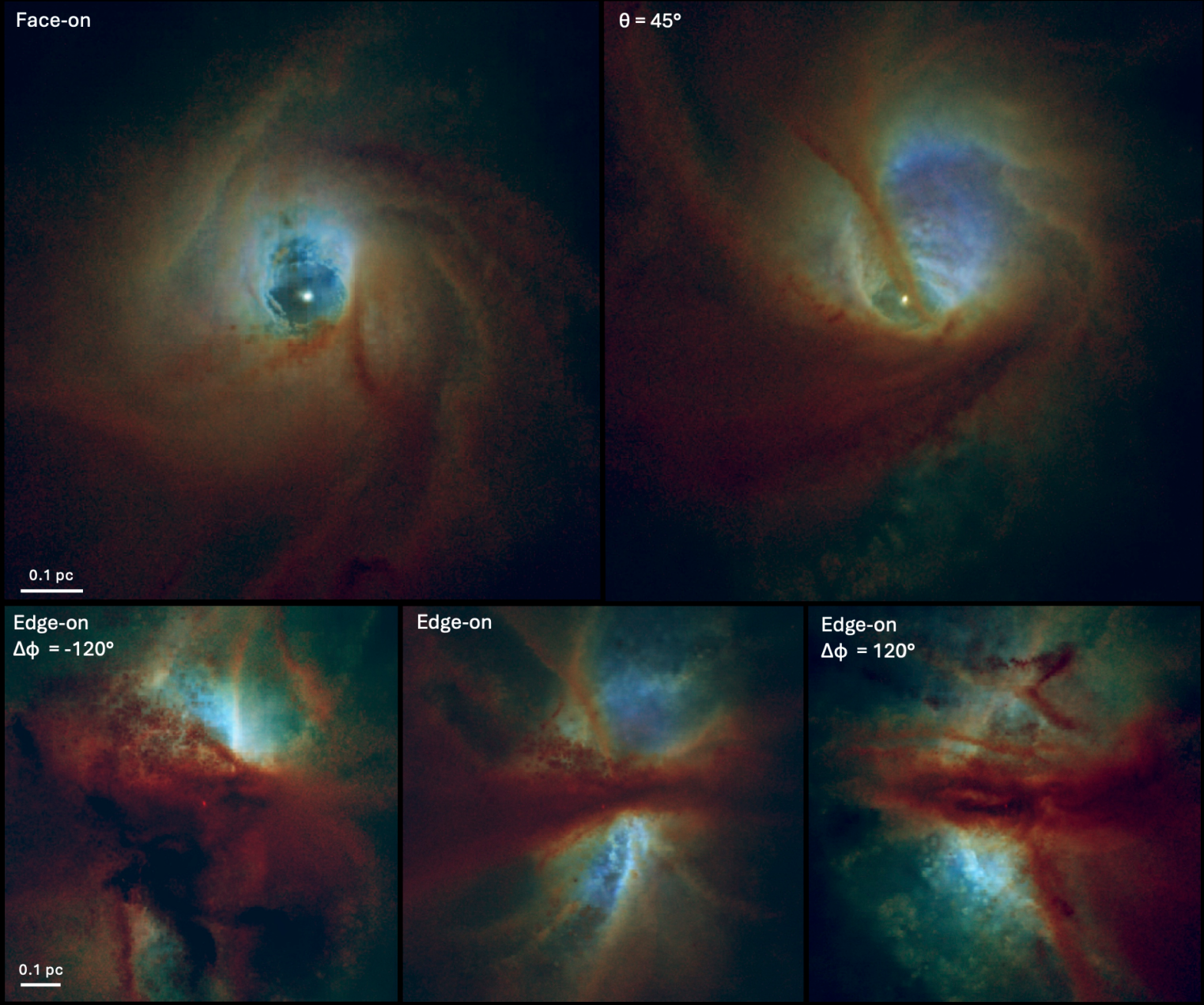}
    \caption{Images from the post-processing radiation emitted out to the black hole radius of influence (BHROI, $\sim$ 1 pc) from the central black hole at five different inclinations and $L_\mathrm{AGN} = 1 \times 10^{44}~\mathrm{erg}~\mathrm{s}^{-1}$, prior to dust self-absorption. The red color indicates emission from wavelengths $\lambda > 37~\mu\mathrm{m}$ (roughly tracing cold dust emission with temperatures $T \lesssim 80~\mathrm{K}$), green indicates emission from $5~\mu\mathrm{m} < \lambda < 37~\mu\mathrm{m}$ ($80~\mathrm{K} \lesssim T \lesssim 600~\mathrm{K}$), and blue indicates emission from $\lambda < 5~\mu\mathrm{m}$ ($600~\mathrm{K} \lesssim T < T_\mathrm{sub} \simeq 1500~\mathrm{K}$). Broadly, the system appears to fit with a standard smooth dust torus model. However, certain orientations reveal significant polar inhomogeneity in the form of clumps and streams as well as significant azimuthal asymmetry. The torus emitting regions correspond best to the highly magnetized outer accretion disk vertically supported by turbulence, in addition to dusty gas tidally torn from the interstellar medium (ISM) complex.}
    \label{fig:RT_1pc_images}
\end{figure*}

\subsection{Radiative Transfer with SKIRT} \label{subsec:skirt}

We use the 3D Monte Carlo radiative transfer (RT) software SKIRT \citep[][]{SKIRT_2020} on FORGE'd in FIRE to perform our post-processing radiation analysis. SKIRT supports multiple anisotropic scattering, absorption and re-emission for a variety of simultaneous media types including dust and electrons, Doppler shifting, polarization by scattering of spherical dust grains or by emission of magnetically-aligned spheroidal dust grains, and self-consistent medium state and radiation field calculation such as with dust destruction and self-absorption.

Due to the complexity of the environment, we implement a sizable portion of SKIRT's current capabilities. In brief, we configure SKIRT to propagate a set of static "primary" radiation sources (e.g. AGN disk or stars) throughout the surrounding media via Monte Carlo radiative transfer, then dynamically update the medium state and emit the resulting dynamic "secondary" media sources (e.g. dust emission, which is determined by the temperature of the dust, a dynamic medium quantity dependent on the radiation field). This process is iterated over until radiative equilibrium is reached, such that the total escaping bolometric luminosity is equal to the input power. Our primary source consists of a central anisotropic AGN accretion disk spectral energy distribution (SED) source. At the scales we study here the stellar contribution is negligible, so we exclude it in this analysis. For our media, we include Compton and Thomson scattering by free electrons, spherical dust grain scattering and absorption for a Milky Way-like dust mix including the 9.7$\mu$m and 18$\mu$m silicate emission/absorption features, dust grain heating, dust grain infrared re-emission, and dust self-absorption. The RT simulation produces spatially-integrated or spatially-resolved SEDs at a variety of inclination angles, along with various dynamic state properties such as the radiation field and the final media temperatures, opacities and densities including the dust properties, stratified by composition and size bins. 

We run SKIRT in 1 pc and 10 pc radius boxes centered around the SMBH to probe the emergent radiation at each scale. These scales were chosen to roughly match the black hole radius of influence for this black hole as a proxy for scales that are distinctly part of the AGN system. In future work, we plan to extend this to distinctly galactic ISM scales ($r \gtrsim 100~\mathrm{pc}$). In our fiducial run, we use a 1 pc radius box around the central SMBH with a \citet{Netzer_1987} profile, central AGN source with $L_\mathrm{AGN} = 5 \times10^{45}~\mathrm{erg}~\mathrm{s}^{-1} \approx 3 L_\mathrm{Edd}$ for $10^3$ isotropically-distributed sightlines plus two face-on directions (polar angles $\theta \in \{0^\circ, 180^\circ\}$) and three edge-on directions (polar angle $\theta = 90^\circ$ and azimuthal angles $\phi \in \{-120^\circ, 0^\circ, 120^\circ\}$), measured relative to the plane of the inner 0.1 pc gas. We show a visualization of a high-resolution, spatially-resolved simulation in a few orientations in Figure \ref{fig:RT_1pc_images}, but leave a discussion of the results to Section \ref{subsec:whatistorus}. 

\subsubsection{Primary Radiation Source}

Since the default FORGE'd in FIRE simulation we study extends down to $\sim 300 R_s$, it does not resolve the full accretion disk and thus a large portion of the bolometric AGN luminosity is unaccounted for. Therefore, we must explicitly model the accretion disk as a source of radiation in our RT simulation. It is commonplace in the literature to approximate the inherent accretion disk emission with a few power-law components since the infrared re-emission does not change significantly as a function of input spectrum shape \citep[e.g.][]{Granato_1994, Schartmann_2005, Honig_2010, Stalevski_2012}. Since we are also interested in the effect of dust absorption to the X-ray to optical/UV spectrum, we also add a soft X-ray and extreme UV component to the model. Our input accretion disk spectrum consists of five power-law components, broadly consistent with \citet{Shen_2020}: X-ray photon index of $\Gamma = 1.9$ up to wavelengths $\lambda = 50~\text{\AA}$, an X-ray to extreme UV value of ext
$\alpha_\mathrm{ux} = -2.4$ (where $f_\nu \propto v^{\alpha}$) up to 500$~\text{\AA}$ followed by an extreme UV component with $\alpha_\mathrm{u} = -1.75$ to $1000~\text{\AA}$ (all together consistent with $\alpha_\mathrm{ox} \sim -1.5$ defined between 2 keV and 2500$~\text{\AA}$), an optical component with $\alpha_\nu = -0.5$ until $\lambda = 1\mu\mathrm{m}$ where it is cut off with a Rayleigh-Jeans tail corresponding to $T \sim 10^4~\mathrm{K}$. We explain our choice of spectrum in more detail in Appendix \ref{app:input_spectrum}.

Since the minimum grid size of our radiative transfer simulations is larger than the $\sim10^{-4}$ pc radius inner accretion boundary in FORGE'd in FIRE, where the majority of the accretion disk emission would lie, we safely make the approximation that this emission is point-like, centered at the SMBH. To account for anisotropic accretion disk emission due to line-of-sight dependent area projection and limb-darkening, we use a standard \cite{Netzer_1987} polar angle distribution 
\begin{equation}
    L_{\lambda}(\theta) \propto 
    \begin{cases} 
      \cos\theta(2\cos\theta + 1) & 0 \leq \theta \leq \frac{\pi}{2} \\
      \cos\theta(2\cos\theta - 1) & \frac{\pi}{2} \leq \theta \leq \pi
   \end{cases}
\end{equation} 
where $\theta$ is measured relative to the angular momentum vector of the inner 0.1 pc, where the circularized accretion disk begins in FORGE'd in FIRE. Due to electron scattering between this primary accretion disk source and the dust sublimation region, this distribution will be reprocessed to roughly match the geometry of the outer, thick accretion disk resolved in FORGE'd in FIRE. The entire SED is normalized such that the bolometric, sightline-averaged luminosity of the primary source is a fixed value $L_\mathrm{AGN}$. Since this simulation assumes radiative equilibrium this also matches the sightline-averaged bolometric luminosity output from the RT simulation. 

While the AGN luminosity can vary significantly on the timescales of the dust torus, we can partially constrain its value. The Eddington luminosity for a black hole of mass $M_\mathrm{BH} = 1.3 \times 10^{7} M_\odot$ is $L_\mathrm{Edd} = 1.6 \times 10^{45}~\mathrm{erg}~\mathrm{s}^{-1}$. \citet{Kaaz_2025} used FORGE'd in FIRE as boundary conditions for the general-relativistic radiation magnetohydrodynamics code H-AMR to extend past the inner accretion boundary down to the ISCO, finding that the SMBH maintains a super-Eddington accretion rate with total system luminosity of $\dot{L}/\dot{L_\mathrm{Edd}} \approx 5 - 10$ for a non-spinning black hole and $\dot{L}/\dot{L_\mathrm{Edd}} \approx 30 - 50$ for a highly-spinning black hole with spin $a = 0.9375$. However, they also report radiation-driven winds/jets which are expected to limit the accretion rate at large radii. Thus, on longer timescales than were simulated by \citet{Kaaz_2025}, this accretion rate is likely an upper-bound. We therefore choose a fiducial value of $L_\mathrm{AGN} = 5 \times 10^{45}~\mathrm{erg}~\mathrm{s}^{-1} \approx 3 L_\mathrm{Edd}$ for most of our runs, but investigate the effect of using different values of $10^{44}~\mathrm{erg}~\mathrm{s}^{-1} \lesssim L_\mathrm{AGN} \lesssim 10^{47}~\mathrm{erg}~\mathrm{s}^{-1}$ (approximately $0.06 L_\mathrm{Edd} \lesssim L_\mathrm{AGN} \lesssim 60 L_\mathrm{Edd}$).

\begin{figure*}[t]
    \centering
    \includegraphics[width=\textwidth]{ 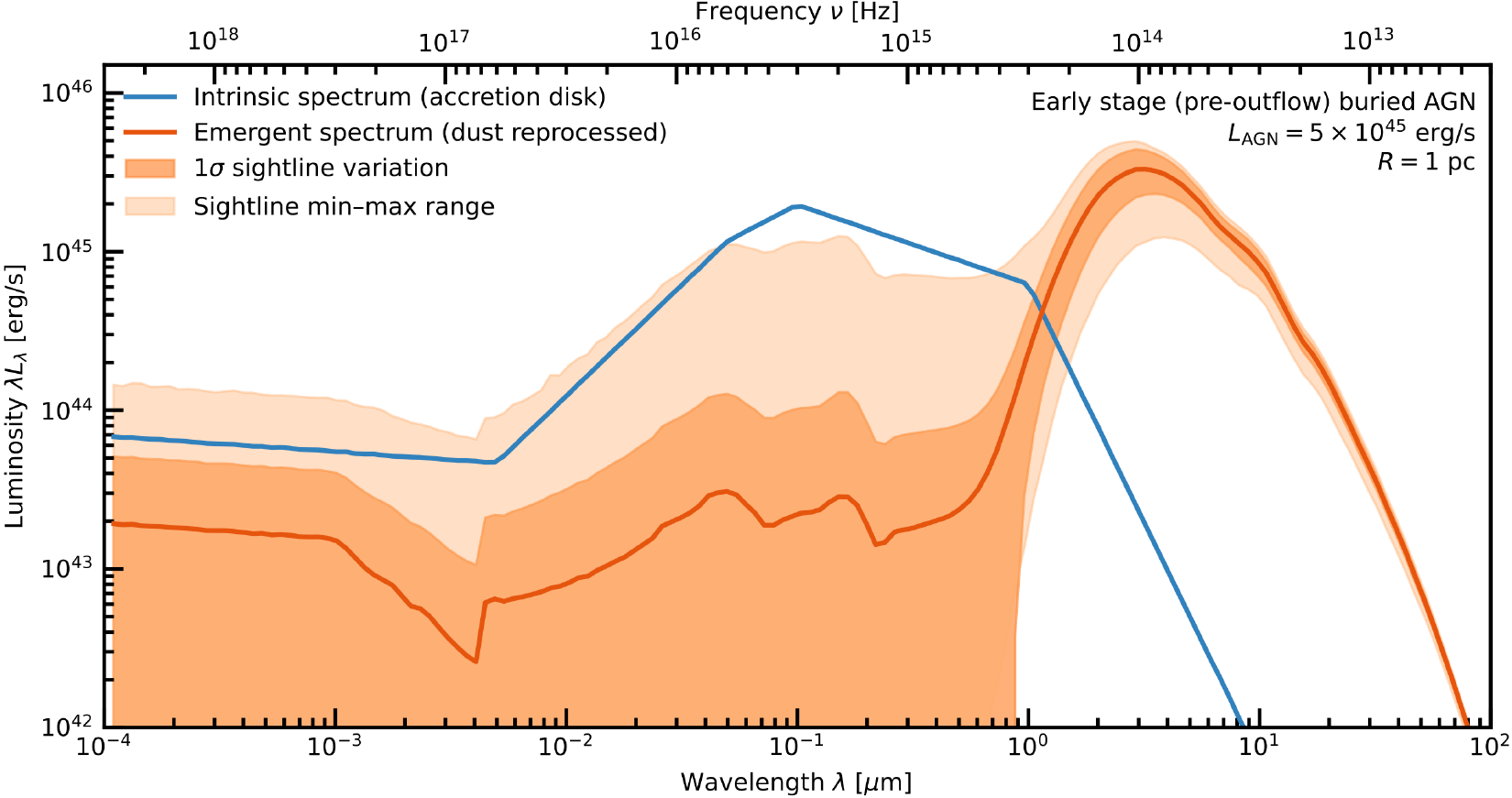}
    \caption{Mean spectra measured at a radius of 1 pc from the central supermassive black hole (SMBH) for an active galactic nucleus (AGN) luminosity of $L_\mathrm{AGN} = 5 \times 10^{45}~\mathrm{erg}~\mathrm{s}^{-1}$. The blue curve shows the intrinsic input spectrum and the orange curve shows the emergent spectrum after reprocessing from Compton scattering and multiple anisotropic dust scattering, absorption and re-emission. The dark orange shading indicates the standard deviation and the lighter shading shows the full range (min-max) of the emergent spectra along our $10^3$ isotropic sightlines. Most emergent sightlines are heavily attenuated in the optical/ultraviolet (UV) and feature a prominent hot dust peak, indicating that this early-stage dusty AGN system is mostly buried. Significant anisotropy is evident even in the mid-infrared, where the spectral luminosity can still vary by a factor of a few.}
    \label{fig:sightline_averaged_spectrum_1pc}
\end{figure*}

\subsubsection{Medium Properties and Secondary Emission}

In order to model dust absorption, scattering and re-emission, we require a model for the dust composition and size distribution. We choose a \citet{Weingartner_2001} Milky Way-like dust mix of graphite, silicate and polyaromatic hydrocarbon (PAH) grain distribution and sizes. In our fiducial run, we split the populations of grains into 9 composition-size bins (3 sizes of graphite, silicate and PAHs each) and allow the density distribution of each bin to change independent of each other due to dust sublimation. We note that splitting the dust into composition-size bins does not significantly affect the simulation output, but it does allow us to roughly study the sublimation structure of the dust as a function of composition and size. We assume an initially even distribution of grains in space (weighted only by size and composition) and allow the simulation to converge the spatial dust grain distribution by iterating over the radiation primary and secondary steps, updating the medium temperatures and sublimating dust accordingly. For silicate dust, we use $T_\mathrm{sub,silicate} = 1200$ K and for graphite dust we use $T_\mathrm{sub,graphite} = 1800$ K, averaging around $T_\mathrm{sub} \sim 1500$ K and consistent with the FORGE'd in FIRE dust mass criterion. To self-consistently destroy the dust according to both the primary (AGN disk emission) and secondary (dust emission) radiation, the primary emission and secondary emission are iterated over. Namely, the radiation field is updated from the primary emission, which is used to update the dynamic media temperatures, opacities and densities, which is then used to produce the secondary emission and updated the radiation field again, repeating this cycle until convergence. This allows us to account for dust self-absorption in our simulation. 

Since part of the accretion disk is resolved in the inner regions of FORGE'd in FIRE, we suspect electron scattering from the outer accretion disk structure may begin to influence our results. We incorporate this in our simulation as a population of free electrons tracing that in FORGE'd in FIRE and interacting with the radiation field via Thomson/Compton scattering. The SKIRT implementation for these processes is discussed in \citet{Meulen_2023}.

\section{Results \& Discussion} \label{sec:results}

\subsection{What is the Torus?} \label{subsec:whatistorus}

It is not simple to determine exactly where the dust torus lies in an ab initio simulation, since there is no clear "doughnut" shape in the dust mass distribution and instead extends out semi-continuously into the surrounding ISM. The dust torus is instead an observational phenomenon, characterized by a region of high optical depth, as viewed from a distance at certain angles, and by having a hot thermal emission component associated to AGN heated dust. This heated dust inner boundary region is easier to define and is clearly set by the sublimation boundary. In the "on-the-fly" FORGE'd in FIRE, which only includes the outer accretion disk emission, the dust sublimates at scales of $\lesssim 0.1$ pc. As we show later, including our fiducial central inner disk component yields inner sublimation radii only slightly larger than this, at $R_\mathrm{sub}\sim 0.25$ pc. The outer boundary of the torus is more difficult to determine since it is not obvious where the "torus" ends and the ISM begins, so we consider an outer bound of approximately the black hole radius of influence (BHROI) or $\sim$ 1-10 pc ($\sim 10^5 R_g$). This is confirmed in our radiative transfer simulations as reproducing the majority of the hot/warm dust emission in this region.

Figure \ref{fig:RT_1pc_images} shows 3-color images of the spatially resolved emergent luminosity from a high-resolution SKIRT simulation with a 1 parsec radius box with $L_{AGN} = 1 \times 10^{44}~\mathrm{erg}~\mathrm{s}^{-1}$ at various inclination angles (described with spherical coordinates $\theta$ and $\phi$). Polar angles $\theta$ are measured from the accretion disk pole estimated as the gas angular momentum vector of a 0.1 pc radius sphere around the black hole. We create these red-green-blue (RGB) images using the emergent flux in the binary filter wavelength bands: $\lambda > 37 \mu\mathrm{m}$ (dominated by warm/cold dust component with $T \lesssim 80 \mathrm{K}$), $5 \mu\mathrm{m} < \lambda < 37 \mu\mathrm{m}$ (hot dust with $80 \mathrm{K} \lesssim T \lesssim 600\mathrm{K}$) and $\lambda < 5\mu\mathrm{m}$ (optical disk emission and very hot dust with $T \gtrsim 600\mathrm{K}$ near its sublimation temperature $T_\mathrm{sub} = 1500\mathrm{K}$), respectively. Note that in these images, we also include various cosmetic effects including arctan scaling, masking the bright central pixel, Gaussian blurring, and image stacking.

While mostly cosmetic, we can still note various torus-identifying properties from Figure \ref{fig:RT_1pc_images}. The bright dot found in the center of many of the images is due to the central AGN disk source and the apparent $\sim$ 0.1 pc-wide "hole" near the center is the boundary of dust sublimation where $T_\mathrm{dust} > 1500\mathrm{K}$. Between the AGN disk and the surrounding dust, there is a region of optical/NIR emission hot dust re-emission as well as scattered disk emission. As seen in the edge-on images, this region extends out to $\sim$0.5 pc perpendicular to the disk, roughly tracing the sublimation boundary. The red regions surrounding the accretion disk indicate a cold/warm dust emission region that strongly obscures the accretion disk emission. Generally, the cold dust is found along the plane of the accretion disk and the warmer dust is mostly in the polar direction. 

\begin{figure}[t]
    \centering
    \includegraphics[width=0.475\textwidth]{ 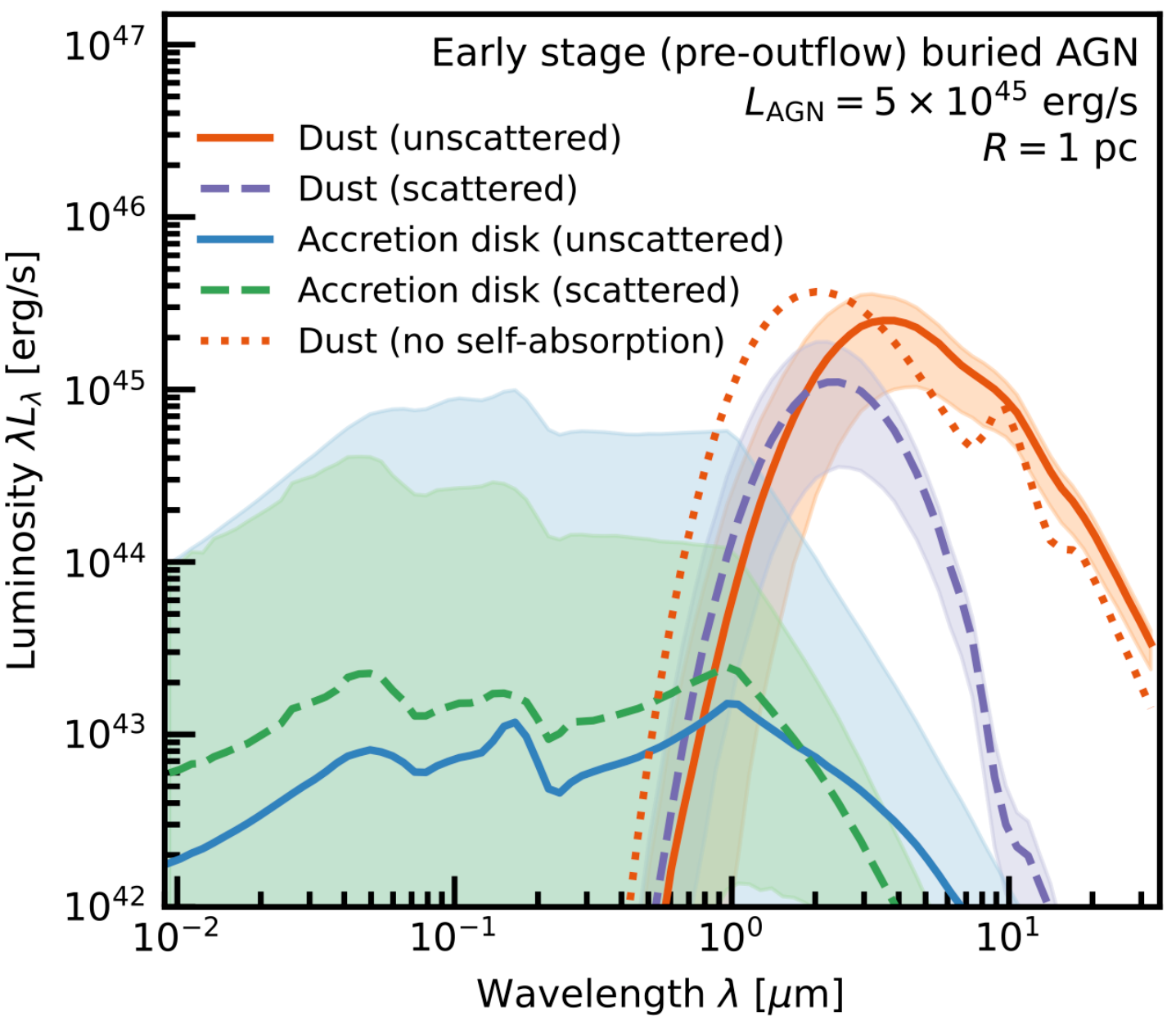}
    \caption{Mean spectra measured at a radius of 1 pc from the central black hole, decomposed into its components with shaded regions showing the sightline (min-max) range. The blue solid line shows the portion of emission originating from the accretion disk that escaped directly to the observer without scattering (only attenuating through absorption) and the dashed green line shows photons that underwent at least one scattering event. The orange solid line corresponds to the thermal dust re-emission that does not undergo a scattering event whereas the purple dashed line corresponds to the dust emission that does. These four curves sum to the total emergent emission. The optical/UV emission is dominated by scattered light. The dotted orange line corresponds to the total dust re-emission if there were no dust self-absorption. Self-absorption both shifts the dust emission towards cooler dust and significantly weakens silicate emission lines.}
    \label{fig:lum_components}
\end{figure}

\begin{figure*}[t]
    \centering
    \includegraphics[width=\textwidth]{ 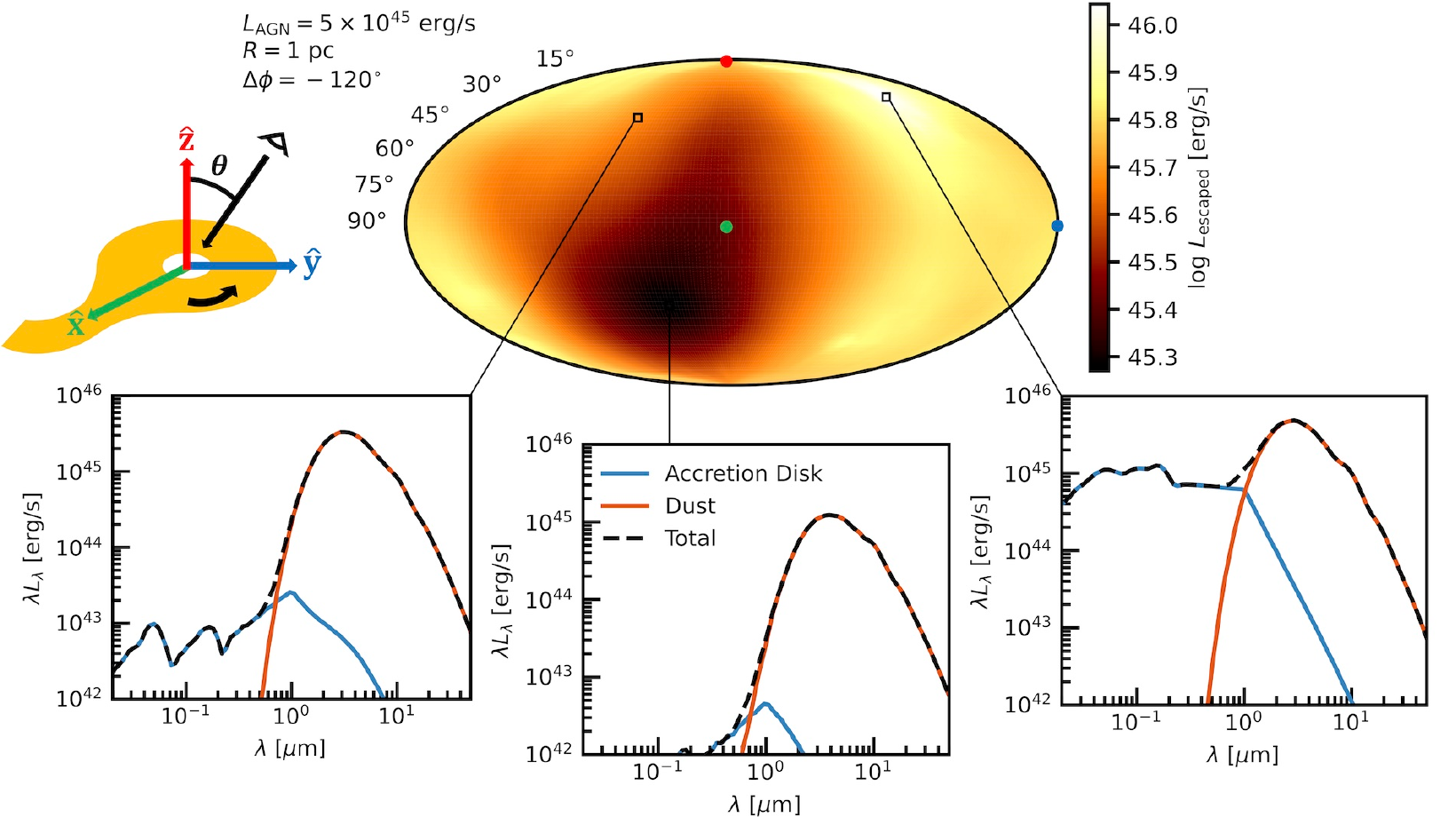}
    \caption{Sightline dependence of escaped bolometric luminosity from post-processing radiative transfer on our fiducial $1$ pc, $L_\mathrm{AGN} = 5 \times 10^{45}~\mathrm{erg}~\mathrm{s}^{-1}$ run, as discussed in Section \ref{subsec:anisotropy}. For illustration purposes, we show Cartesian axes aligned such that the angular momentum of the outer accretion disk lies along the z axis, the inflowing stream of dusty gas lies roughly along the x axis, and the polar angle $\theta$ is measured from the z axis. We display the escaping bolometric luminosity as a function of sightline using a Mollweide projection. Most of the extinction in the escaped luminosity is dominated by a dusty stream of cold inflowing gas. We also plot the maximum, minimum and mean $L_\mathrm{bol}$ sightline spectra, showing that both the optical and IR change as a function of sightline.}
    \label{fig:lum_angle_dependence_1pc}
\end{figure*}

Figure \ref{fig:RT_1pc_images} indicates a dust structure that tends to match broadly simple analytical torus models, but has significant polar anisotropy and azimuthal variation. Matching with standard expectations, there is an increase of dust extinction roughly along the accretion disk plane, which heavily obscures direct emission from the disk if aligned near edge-on. However, the effective cold dust scale height $H/R$ varies from $\sim$0.2 -- 0.6. There thus appear to be two components of the torus: 1) the standard, mostly azimuthally symmetric outer accretion disk region, and 2) a stream of cold, high-density dust that fuels the accretion disk and causes azimuthal asymmetry. There is also some evidence that part of this structure is composed of self-gravitating dust clumps, particularly in polar regions. Some of the polar anisotropy also comes from independent polar dust streams. Both streams and clumps shield the dust behind it from central radiation, leading to colder dust nearer to the accretion disk than would be otherwise. 

The dust torus is therefore primarily composed of the outer accretion disk in addition to some dusty gas, which has been tidally torn away from the surrounding ISM complex contributing to azimuthal anisotropy in the dust structure. In the FORGE'd in FIRE simulation, this region of the accretion disk has Toomre $Q \gtrsim 1$, scale height $H/R \sim \delta v_z/v_\mathrm{circ} \sim 0.3 - 0.8$, sonic Mach number $\mathcal{M}_s \equiv \delta v_\mathrm{turb}/c_s \sim 10 - 60$, and Alfvénic Mach number $\mathcal{M}_A \equiv \delta v_\mathrm{turb}/v_A \sim 1$ (so plasma $\beta = c_s^2/v_A^2 \ll 1$), corresponding well to the primarily torodial magnetic field pressure-dominated region of the quasi-steady state super-Eddington (with $\dot{M} \sim 10-100~M_\odot$/yr) accreting disk with little to no star formation or fragmentation. \citet{Hopkins_2024b} show in detail the properties of this system and how it arises naturally from the advection of magnetic flux from the ISM. They also show that simulating without magnetic fields leads to catastrophic fragmentation and a much smaller disk. This dust structure is therefore also likely vertically supported by the same trans-Alfvénic super-sonic turbulent support characteristic of the magnetically-dominated accretion disk model \citep{Hopkins_2024d} which naturally arose in FORGE'd in FIRE. These strong magnetic fields are crucial in supporting not only the outer accretion disk, but also the dust torus structure in FORGE'd in FIRE. Moreover, at these early stages in the simulation, there is distinctly no dusty wind ($\dot{M}_\mathrm{in} \gtrsim \dot{M}_\mathrm{out}$). Of course, it is possible that at later stages a dusty wind can be created from AGN feedback in the form of outflows originating from the central SMBH. We explore this possibility and its effects on our results in Section \ref{subsec:outflow_analysis}.

\begin{figure*}[t]
    \centering
    \includegraphics[width=\textwidth]{ 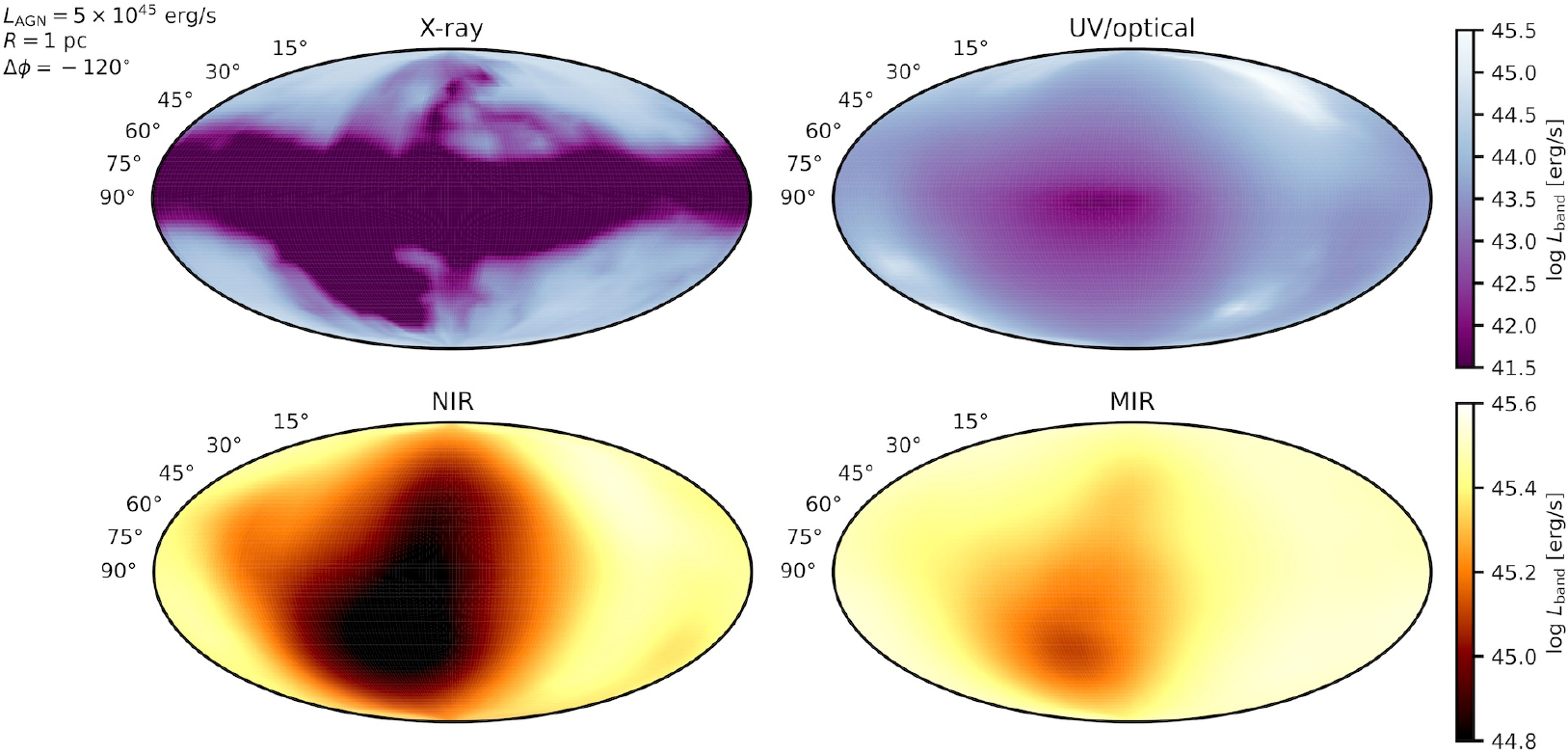}
    \caption{Sightline dependence of escaped luminosity in the fiducial run as in Figure \ref{fig:lum_angle_dependence_1pc}, but now separated into wavelength bands. We define here the X-ray emission as $\lambda < 100~\text{\AA}$, the UV/optical as $0.01~\mu\mathrm{m} < \lambda < 0.8~\mu\mathrm{m}$, the near infrared (NIR) as $0.8~\mu\mathrm{m} < \lambda < 3~\mu\mathrm{m}$, the mid infrared (MIR) as $3~\mu\mathrm{m} < \lambda < 25~\mu\mathrm{m}$. What little emission escapes in the X-ray and UV/optical is highly anisotropic due to significant dust extinction and Compton scattering. The NIR and MIR, which dominate the escaping emission at this scale and primarily originate from self-absorbed dust emission from the hot/warm dust near the SMBH, but can still vary by a factor of a few due to reprocessing from the dusty stream fueling accretion.}
    \label{fig:lum_band_dependence}
\end{figure*}

\subsection{Early Stages: A Buried AGN} \label{subsec:spectra}

In Figure \ref{fig:sightline_averaged_spectrum_1pc}, we show the mean (sightline-averaged) emergent spectra to 1 pc from the SMBH, for a fiducial AGN disk luminosity $L_\mathrm{AGN} = 5\times 10^{45}~\mathrm{erg}~\mathrm{s}^{-1}$. We also show the $1\sigma$ (standard deviation) variation and min-max range across sightlines in addition to the mean intrinsic spectrum corresponding to the input accretion disk source prior to any dust reprocessing. 

The IR emission is very hot, peaking at a wavelength of $\sim$ 3-4$~\mu\mathrm{m}$. Due to our boundary cut, we do not see a large cold dust component, however this is expected to appear at larger distances from the black hole, as the radiation is reprocessed by the surrounding cold ISM dust. It is not too surprising to find some sightline anisotropy in the NIR, on the order of $~1$ dex. However, this anisotropy continues well into the MIR, varying by a factor of $\sim 4$ at around $10~\mu\mathrm{m}$. We explore the source of this anisotropy in Section \ref{subsec:anisotropy}.

The optical/UV component is heavily extincted by the dust along the majority of the sightlines. The mean $\lambda L_\lambda$ dust reprocessed spectrum in the optical is $\sim$2 dex lower than the accretion disk input component and $\sim$3 dex lower than the IR component. However, there is some optical emission in a small number of sightlines, as evidenced by the large range in optical/UV emission. The X-ray emission is also heavily extincted. Indeed, the gas from FORGE'd in FIRE in this region has column densities of $n_H \sim 10^{24-26}~\mathrm{cm}^{-2}$ and is thus Compton-thick.

Figure \ref{fig:lum_components} decomposes the mean spectra into the scattered and unscattered emission originally sourced from the accretion disk or the dust. It also overlays the total dust emission curve when rerunning explicitly without dust self-absorption (i.e. re-absorption of the hot IR component). It is clear that there is significant self-absorption since the dust without self-absorption is significantly redder than with, which is consistent with the optically thick medium picture of this system.

Our dust mix includes silicate features relevant to the dust torus, which are often used as a parameter to distinguish between dust torus models. In our mean spectrum, there is a very weak presence of 9.7 $\mu$m and 18 $\mu$m silicate emission features in our fiducial mean spectrum. We also notice that the silicate emission line, which is strong without self-absorption, is highly suppressed when including self-absorption, further supporting that this is a buried AGN.

Along orientations that do have an optical/UV component, the scattered optical/UV component is higher than the unscattered case. This indicates that the small number of orientations that appear type 1-like ($\sim$5\%) possibly escape via thin channels where there are multiple scattering events. 

Overall, in these early (pre-feedback) stages, this system looks like a buried, Compton-thick AGN with only a small fraction ($\sim$5\%) of sightlines showing optical/UV emission. Indeed, this is not unexpected for a major galaxy merger such as this one, and suggests that this system may appear or evolve to appear as a warm ultra-luminous IR galaxy (warm ULIRG) or hot dust obscured galaxy (hot DOG) on a galactic scale. We plan to extend this work to larger distances from the black hole, including the stellar and ISM dust contribution of this particular galaxy in the future.

\begin{figure*}[t]
    \centering
    \includegraphics[width=\textwidth]{ 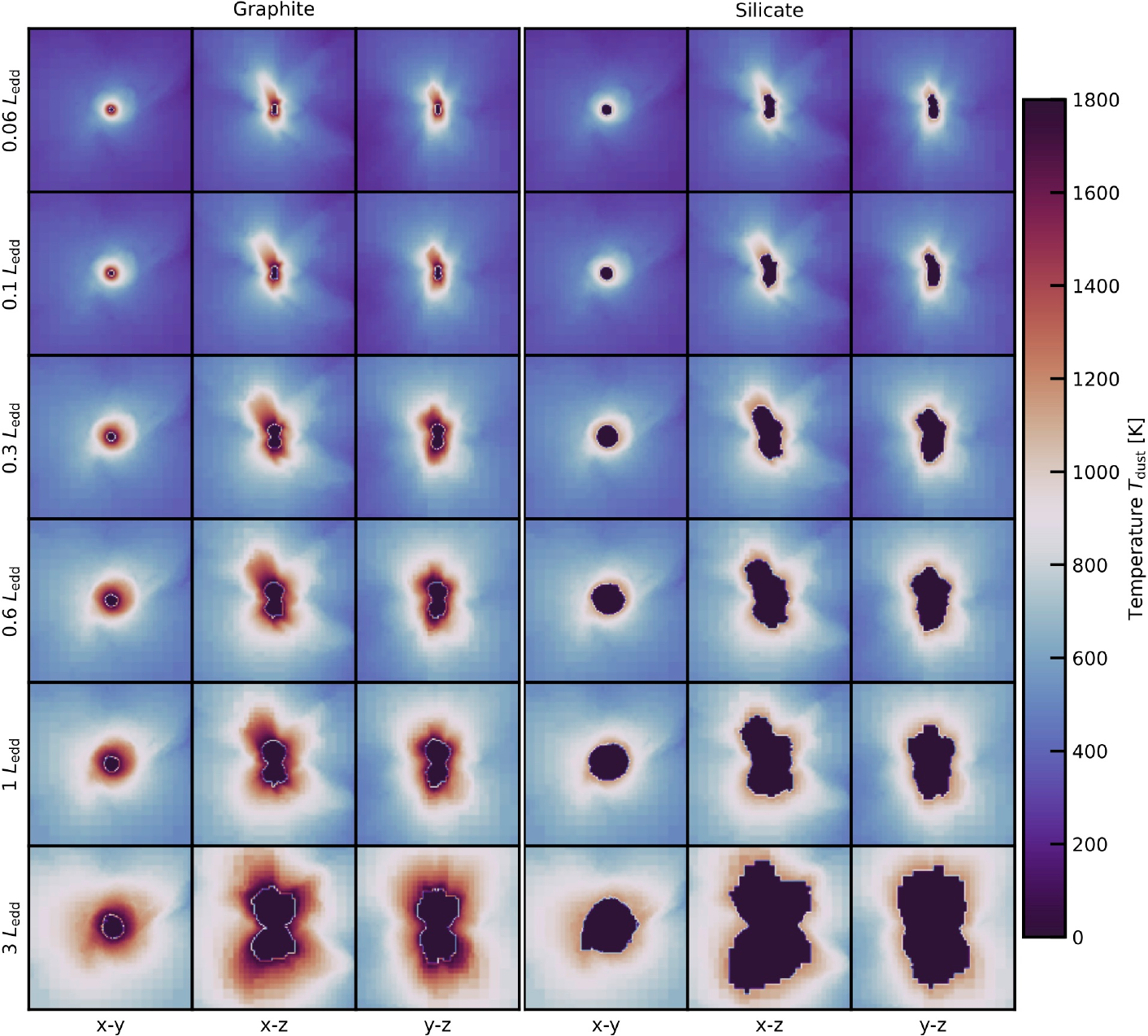}
    \caption{Temperature axis slice plots for silicate and graphite grains showing the sublimation structure ($T_\mathrm{sub, silicate} = 1200~\mathrm{K}$ and $T_\mathrm{sub, graphite} = 1800~\mathrm{K}$) in black, surrounded by a white ring. The sublimation radius grows as a function of luminosity and is stratified across grain composition and size bins. The polar angle sublimation structure distribution is roughly the same as the \citet{Netzer_1987} accretion disk distribution, particularly in graphite and for low $L_\mathrm{AGN}$, but acquires significant anisotropy at larger $L_\mathrm{AGN}$. The average sublimation radius also roughly follows the standard $L_\mathrm{AGN}^{0.5}$ trend.}
    \label{fig:sublimation_structure_by_comp}
\end{figure*}

\begin{figure*}[t]
    \centering
    \includegraphics[width=\textwidth]{ 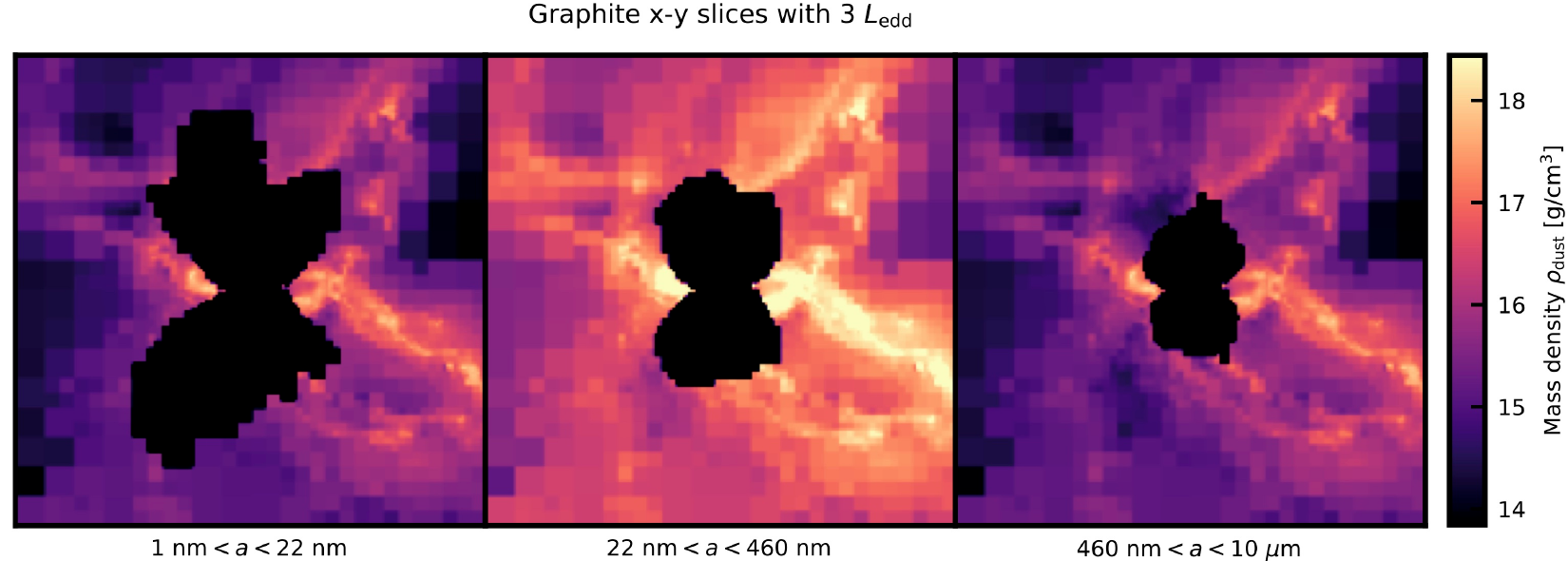}
    \caption{Dust mass density x-z axis slice plots for $L_\mathrm{AGN} = 3 L_\mathrm{edd}$ as a function of graphite dust grain size bin, showing the sublimation region in black. Smaller grains tend to sublimate easier, leading to dust size stratification with a concentration of large dust grains in the innermost regions. The inhomogeneity of the accretion disk and surrounding polar dust streams/clumps causes anisotropy in the sublimation structure by shielding certain outer regions from heating by the central source.}
    \label{fig:substruct_by_size}
\end{figure*}

\subsection{Anisotropy of the MIR} \label{subsec:anisotropy}

In this section, we investigate the origin of the MIR anisotropy. It is known that many galaxies exhibit a polar extended mid-IR component thought to come from radiation-driven dusty winds \citep{Honig_2012, Honig_2013, Lopez_Gonzaga_2016, Garcia_Bernete_2022}. Since there are no dynamics simulated by our post-processing radiation simulations, this is distinctly not what is observed here. However, we still find significant anisotropy in the MIR in Figures \ref{fig:sightline_averaged_spectrum_1pc} and \ref{fig:lum_components}. It is not immediately clear where this anisotropy originates. Figure \ref{fig:lum_components} gives a clue to the potential source, as the anisotropy is solely due to absorption effects and not scattering, suggesting that the asymmetry may arise directly from the geometry of the optically thick medium. First, we investigate the anisotropy of sightline-inferred bolometric luminosity.

Figure \ref{fig:lum_angle_dependence_1pc} shows the total "escaped" luminosity as a function of angle. This is the bolometric luminosity as inferred from a given direction (i.e. $L_\mathrm{escaped}(\Omega) = F(\Omega)/4\pi d^2$, where $F(\Omega)$ is the flux observed from a distance $d$ to the system in a given direction). The sightline mean of the escaped luminosity is thus necessarily the input accretion disk luminosity $L_\mathrm{AGN}$ since we reach radiative equilibrium. In Figure \ref{fig:lum_angle_dependence_1pc}, we also depict the location of the outer accretion disk and the dusty stream of gas that fuels the accretion disk and originally broke away from the local ISM cloud. 

The escaped luminosity anisotropy can vary by almost an order of magnitude and is spatially correlated. Since our system is mostly buried, the bolometric luminosity is dominated by the NIR/MIR components. The majority of this anisotropy of the manifests as azimuthal asymmetry, rather than polar asymmetry. In particular our system exhibits a "dark side" where there is less bolometric emission. This dark side is directly spatially correlated with the direction of the dusty inflowing gas, broken away from the local ISM cloud and fueling accretion. The luminosity is minimum in the direction of the dusty stream, but is maximized in a mostly polar direction offset from the axis of the stream. 

The spectra associated to the sightlines with minimum, maximum and mean $L_\mathrm{escaped}$ are also shown in Figure \ref{fig:lum_angle_dependence_1pc}. In the direction of most extinction, there is little to no optical/UV emission as well as a hot dust peak that is suppressed compared to the mean. In the direction of least extinction, there is little to no deviation in the hot dust peak from the mean spectrum, but there is significant escaped optical/UV emission. This suggests that there are two factors that influence the anisotropy of the escaping luminosity: the amount of escaped optical/UV emission and the amount of NIR/MIR extinction. 

Showing the inferred band-convolved escaping luminosity in each direction (Figure \ref{fig:lum_band_dependence}), the source of the anisotropy becomes clearer. In the X-ray, most of the emission escapes except along the mid-plane roughly according to the intrinsic accretion disk \citet{Netzer_1987} polar anisotropy, as well as some regions of particularly high Compton thickness corresponding to polar substructure including thick streams and clumps. However, the X-ray luminosity is too low to drastically influence the bolometric escaped luminosity anisotropy. In the optical/UV, there are only a few directions that allow radiation to escape since the system is mostly buried. These directions tend to lie in the polar region and correspond to directions away from the fueling gas stream axis with particularly low dust density and where the sublimation region is more extended (as shown in Section \ref{subsec:substructure}). In the NIR and MIR, the polar dependence begins to disappear, but the azimuthal asymmetry becomes clear. It is commonly assumed that these wavelength regimes are nearly isotropic, exhibiting only some NIR extinction in particularly optically thick regions. Traditionally, these regions are in the direction of the "torus" and are azimuthally symmetric. In principle, however, there is no reason that dust tori require the azimuthally symmetric systems assumed in many phenomenological torus models. In this system, the dusty stream is so dense that it dominates the anisotropy in the NIR and even retains significant anisotropy in the MIR. 

Thus, an emergent picture arises, where there are two distinct parts of the dust torus: 1) the uniform, hot dust emitting outer accretion disk region and 2) the cold, dense stream of dusty gas fueling accretion that obscures NIR/MIR emission. This second NIR/MIR obscuring region is the source of azimuthal asymmetry. Our system also happens to be buried except along a few, scattering-dominated sightlines leading to certain channels of optical/UV emission, which contributes some, mostly polar, anisotropy. 

\begin{figure}[t]
    \centering
    \includegraphics[width=0.46\textwidth]{ 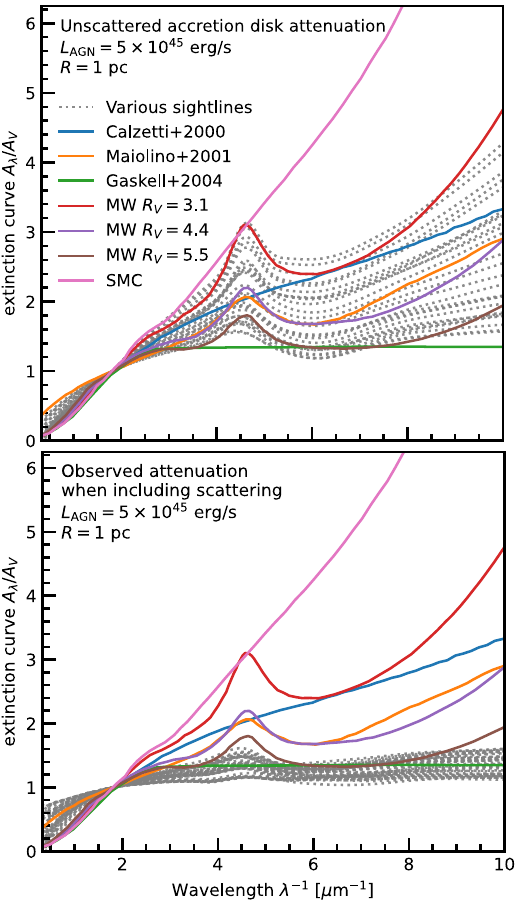}
    \caption{Extinction curves of the output as a function of a few sightlines, defined as $A_\lambda = -2.5\log_{10}(F_\lambda/F_\mathrm{AGN, \lambda})$, where $F_\mathrm{AGN, \lambda}$ is the mean spectral flux density of the central disk source if there were no intervening medium and $F_\lambda$ includes the medium. The top plot only includes emission that has only been absorbed, whereas the bottom plot also includes emission that has undergone scattering. In each plot, we show a few sightlines in gray dotted lines and a few comparison extinction curves in dashed lines, including \citet{Calzetti_2000}, \citet{Maiolino_2001a} and \citet{Gaskell_2004}. The unscattered extinction curve demonstrates the affect that sublimation-driven stratification of dust grain composition and size distribution has on the originally Milky-Way (MW) dust composition/distribution. The mean is roughly MW with $R_v=4.4$ or \citet{Maiolino_2001a}-like, but vary significantly in shape across sightline. However, when including the scattered emission, the extinction curves significantly grayed similar to \citet{Gaskell_2004}, since a large portion of the emission is scattered due to the high optical depth of the system.}
    \label{fig:extinction_curves}
\end{figure}

\subsection{Stratified Sublimation Structure} \label{subsec:substructure}

Since we include a dust mix with 9 bins of grain composition and size in all of our simulations, we can investigate the sublimation structure as a function of dust composition/size and AGN luminosity. Figure \ref{fig:sublimation_structure_by_comp} shows the dust temperature map for graphite and silicate bins in slices for all three direction planes where x-y the plane of the accretion disk for $L_\mathrm{AGN} = 0.06$-$3 L_\mathrm{Edd}~\mathrm{erg}~\mathrm{s}^{-1}$. Figure \ref{fig:substruct_by_size} shows the mass density for the graphite bins as a function of grain size. There are differences in the sublimation structure across composition/size that ultimately originate from a combination of the geometry of the system, the sublimation temperature (here, set as $T_\mathrm{sub, graphite} = 1800$ K, $T_\mathrm{sub, silicate} = 1200$ K) and the relative opacities between dust grain species/sizes. 

Sublimation structure is generally anisotropic, partly due to intrinsic \citet{Netzer_1987} polar anisotropy from the accretion disk spectrum, but also to small clumps or streams of dust that obscure regions of dust from the central source in certain polar directions. This shielding effect of clumps and streams causes cold dust to exist near the black hole more than it would otherwise, particularly at larger $L_\mathrm{AGN}/L_\mathrm{edd}$. The sublimation structure also broadly follows the standard $R_\mathrm{sub} L^\alpha$ with $\alpha\sim0.5$ power law expected. It also recovers the trend of slope $\alpha$ at higher observed effective wavelengths and is surprisingly consistent with the observational dependence of $\alpha$ as a function wavelength band used (broadly, $\alpha$ decreases with effective wavelength of the band).

\subsection{Grayed Extinction Curves} \label{subsec:extcurves}

We might expect the stratification of the sublimation structure to affect the inferred extinction curves. Figure \ref{fig:extinction_curves} shows the extinction curves of our fiducial run defined as $A_\lambda = 2.5\log_{10}(F_\lambda/F_\mathrm{AGN, \lambda})$ with $F_\mathrm{AGN, \lambda}$ being the intrinsic AGN accretion disk spectrum and $F_\lambda$ is the emergent specific flux after including the medium. 

In the top panel of Figure \ref{fig:extinction_curves}, we show the extinction curves of the escaped, unscattered emission to isolate the effect of stratification alone. Since we initially assume a MW dust composition, any deviation from this shape in the unscattered emission is a result of the sublimation dust grain size and composition stratification. The mean unscattered extinction curve is very similar to the galactic $R_V = 4.4$ and \citet{Maiolino_2001a} curves, but can vary across observed sightlines from the input $R_V = 3.1$ to $R_V \sim 5.5$. Stratification thus generally grays the spectrum since the dust grains are preferentially larger within the stratification zone (as shown in Section \ref{subsec:substructure}). 

In the observed spectrum, including scattered emission (bottom panel of Figure \ref{fig:extinction_curves}), the extinction curve is very flat and similar to the "nuclear" extinction curves from the composite spectra of \citet{Gaskell_2004}. Since most of the optical/UV emission escapes through small channels of scattering-dominated emission (as shown in Figures \ref{fig:lum_components} and \ref{fig:lum_band_dependence}), the input spectrum is effectively mirrored in through these channels, producing a very flat extinction curve. Thus, the sublimation structure stratification can lead to a plethora of extinction curves on its own, but it is difficult to notice in this particular system, since it is a highly optically thick system. 

Although gray spectra such as \citet{Gaskell_2004} are commonly found by composite spectral methods, such methods are often disfavored because they are argued to be less accurate than individual reddened quasar observations \citep[e.g. arguments by][]{Willott_2005}. Indeed, many individual reddening observations, such as the analysis of SDSS data in \citet{Krawczyk_2015}, find instead that the observed extinction is MW or SMC-like and thus likely more concentrated with small dust grains. We stress that this is not entirely inconsistent with our findings for a few reasons. Firstly, studies using individually lightly reddened quasars necessarily use systems that appear type 1-like and have significant optical/UV emission, whereas this system is mostly buried. Even in the few directions where optical/UV does manage to escape, the AGN optical/UV is heavily obscured and would likely not appear in such a sample. If we are correct in arguing that this system will later transform into a standard type 1/2 (as discussed in Section \ref{subsec:outflow_analysis}), then the outflow dust thermochemistry will determine the system's true extinction curves in its optical/UV-observable stage. Moreover, we currently only measure "nuclear" regions within the BHROI, argued to have gray extinction by \citet{Gaskell_2004} because of effects such as the sublimation of small grains found here. It is therefore possible that, while the majority of the hot dust emission lies within the BHROI, the galactic nucleus ISM extinction still determines a large portion of the observed optical/UV extinction curve. In such a case, the extinction would appear closer to the surrounding ISM extinction, rather than the nuclear regions, which include graying effects such as sublimation. In future work, we plan on extending our analysis here to galactic scales to answer such questions in this particular system. Finally, we assume an initial MW mean dust composition and size distribution, which causes the unscattered light to naturally appear MW-like and $R_V \geq 3.1$ before any sublimation effects which gray the curve further. It is possible that some effect beyond the scattering geometry and sublimation structure, such as dust shattering, could cause the local environment to be preferentially small grain dominated.

\begin{figure}[t]
    \centering
    \includegraphics[width=0.46\textwidth]{ 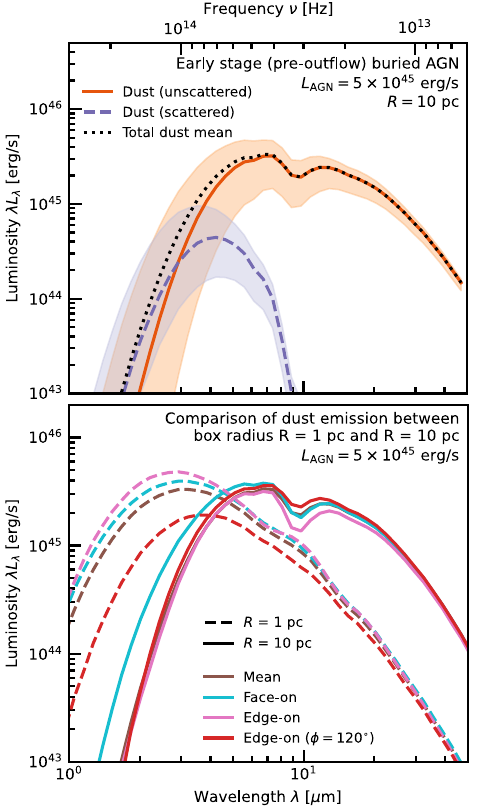}
    \caption{Dust emission spectra including all dust within a radius of 10 pc, as discussed in Section \ref{subsec:extendingout}. The top plot shows the mean spectra and corresponding sightline (min-max) range, separated into the scattered and unscattered dust emission. The bottom plot shows a side by side comparison of the $R=10~\mathrm{pc}$ spectrum and the $R=1$pc spectrum in a few sightlines. The infrared anisotropy remains present into the surrounding dust and some of the hot dust is self-absorbed and re-emitted by the colder surrounding dust. We also notice that the weak silicate emission features as observed from $R = 1~\mathrm{pc}$ become absorption features at $R = 10~\mathrm{pc}$. }
    \label{fig:spectra_10pc}
\end{figure}

\subsection{Extending into the ambient ISM} \label{subsec:extendingout}

We now briefly extend our analysis into regions nearing the ambient ISM to determine differences after local reprocessing. Figure \ref{fig:spectra_10pc} shows the infrared spectra of the system including all material in a 10 pc radius box for $L_\mathrm{AGN} = 5 \times 10^{45}~\mathrm{erg}~\mathrm{s}^{-1}$, and its comparison to our fiducial 1 pc radius box simulation. The dust shifts to lower wavelengths as the spectra begin to be reprocessed by the colder surrounding dust, now peaking closer to $\sim 6-7~\mu\mathrm{m}$. Moreover, dust absorption causes the very weak emission silicate features to become absorption features. These silicate absorption features are often seen in observations of deeply obscured AGN systems \citep{Spoon_2007, PerezBeaupuits_2011, GarciaBernete2022}. Many of the other emergent properties continue to remain present, including MIR anisotropy. We defer a detailed analysis to future work, where we plan on extending our radiative analysis well into the galactic nucleus to better understand this system in the context of its environment.

\subsection{Late Stages: Bipolar Outflow Model}  \label{subsec:outflow_analysis}

Up to this point in our analysis, it is clear that our system at its current stage appears buried beneath optically thick dust. While this is not unexpected for galaxy merger systems, not all of these systems are type 2. Moreover, similar systems have been observationally shown to often contain molecular outflows. Were there such an outflow to occur, it may clear out a path so that the system would look type 1-like along certain sightlines \citep[e.g.][]{Hopkins_2016}. As it turns out, recent extensions of FORGE'd in FIRE to ISCO in \citet{Kaaz_2025} and \citet{Hopkins_2025} show that after initially starting accretion, there is a fast outflow that naively has enough kinetic energy to reach and potentially surpass dust torus scales. Motivated by this, in this section we rerun our fiducial RT simulation earlier, but now on initial dust density distribution from FORGE'd in FIRE that removes a bicone to model a possible outflow. In particular, we decrease the density in a bicone cutout by a factor of $10^3$ perpendicular to the accretion disk plane and with an opening angle of $30^\circ$, which are roughly consistent with the results from these further zoom-ins.

\begin{figure}[t]
    \centering
    \includegraphics[width=0.45\textwidth]{ 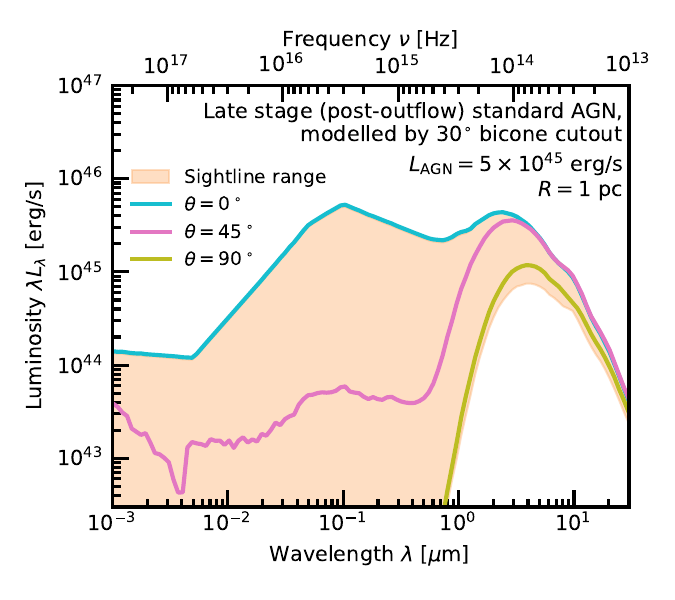}
    \caption{Spectra of the fiducial run after cutting out a 30 degree bicone, modeling the "late-stage" effects of AGN feedback that clears an optically thin path. We show the emergent spectrum face-on to the system in cyan ($\theta = 0^\circ$), an edge-on spectrum in green-yellow ($\theta = 90^\circ$), and an oblique angle in pink ($\theta = 45^\circ$). The "late-stage" spectrum is now optically thin in the face-on directions but still optically thick in the edge-on directions. More optical/UV emission is observed in the oblique, non-cutout directions than before primarily due to scattering from emission off the bicone region. Outflows that burrow an optically thin path in the polar can thus transform our buried system into a traditional type 1/2 system. Many of the properties of the original system remain, including IR anisotropy and silicate emission/absorption features. In fact, the NIR and MIR anisotropy is further enhanced due to a shortage of edge-on infrared emission compared to the non-cutout system caused by previously trapped light in the mostly buried system, which can now escape in the polar direction.
    }
    \label{fig:sightline_averaged_bicone_cutout_spectrum_1pc}
\end{figure}

Figure \ref{fig:sightline_averaged_bicone_cutout_spectrum_1pc} shows the resulting spectra of our bicone cutout model. Of course, now the system appears as a typical type 1/2 system in the optical/UV, since sightlines with polar angle $\theta \leq 30^\circ$ are optically thin and directly return the input accretion disk spectrum. Along oblique angles, outside the cutout, more optical/UV leaks from the system than before the cutout. This optical/UV oblique angle emission is due to light that scatters from the bicone region and dominates the unscattered emission directly from the accretion disk and the scattered emission within the torus itself. There is still a hot dust component with a similar shape to that observed in the fiducial, non-cutout run. The IR normalization is lower than in the fiducial case, since a significant amount of the emission that would otherwise be trapped in the buried system is now allowed to escape in the polar region. The IR emission is still highly anisotropic and is spatially correlated in the same way as the fiducial spectra. In fact, the lower IR normalization in the edge-on spectra slightly enhances this anisotropy in the cutout model spectra. Other IR-dominated properties are generally also retained, such as the weak silicate emission/absorption features. Thus, we argue that many of the properties discussed here may remain after a self-consistent analysis of the "late-stage" (post-feedback) dust torus state.

\subsection{General Caveats}\label{subsec:caveats}

This work is accurate to the early-stages of the dust torus formation in our system, and may change at "late-stages" after AGN feedback in the form of radiatively accelerated winds or SMBH outflows. The first FORGE'd in FIRE run accurately tracks the formation of the quasar within resolved radii, but did not include any AGN feedback prescriptions beyond treating the $\sim 10^{-4}$ pc inner boundary as a purely inflowing sink. Thus, this quasi-steady state dust torus could evolve its state after AGN radiation or feedback at later stages. A preliminary analysis using the first-order effect of radiative pressure from SKIRT suggests that including radiation feedback from the black hole may induce a polar radiatively-driven dusty outflow, which could drive a hole through the buried system and allowing optical/UV emission escape. Moreover, extensions of FORGE'd in FIRE down to ISCO \citep{Kaaz_2025, Hopkins_2025} show that there is a significant outflow from the SMBH that could plausibly reach and influence dust torus regions after $\sim 10^4-10^5$ years. While we speculate on the effect of such feedback in Section \ref{subsec:outflow_analysis}, a completely self-consistent inclusion of AGN feedback would require explicitly tracking the dynamics of the outflow and radiative feedback by progressively zooming-out until we reach the torus region. The challenge of such a simulation lies in accurately integrating the central wind/jet launching region, where the orbital time is on the order of seconds, to the very long dust torus region timescales. We plan on pursuing this in future work. 

Although most of the system is buried, we do see evidence of optical/UV emission escaping through small, scattering-dominated channels. However, in lower spatial resolution SKIRT runs, including our 10 pc fiducial run, many of these effects disappear as they are averaged into larger spatial grid bins. This is a well-known limitation of radiation transport on grids and can affect the resulting spectra of systems that are clumpy or have small escape channels on small scales. In the same way, due to the spatial resolution limit of both FORGE'd in FIRE and of our SKIRT runs, we may not resolve some substructure vital to the observed radiation. It is therefore possible that a non-negligible amount of primary source emission from the accretion disk could escape further through the dust, potentially altering the resulting spectra. This would not drastically affect the sublimation boundary analysis, since the majority of the input radiation is only weakly affected by Compton scattering between the SMBH and the sublimation radius. However, the IR spectra, which are governed by self-absorption and the extinction curves could change due to a lack of substructure. Although it is generally thought that there is a limit to the number of clumps before it produces unrealistic spectra \citep{Nenkova_2008}, some amount of substructure in the form of clumping or optically thin channels could allow more optical/UV to escape, potentially reducing some of the hot dust emission and IR anisotropy observed.

\section{Conclusions} \label{sec:conclusion}

FORGE'd in FIRE \citep{Hopkins_2024a} was the first multi-scale ab initio simulation of the formation of a quasar, starting from cosmological initial conditions and zooming to $\sim 300$ Schwarzschild radii from the central $M_\mathrm{BH} = 1.3 \times 10^7 M_\odot$ black hole. FORGE'd in FIRE contains a wide range of important physics at these scales including adaptive multi-band radiation hydrodynamics, coupled to the non-LTE and non-equilibrium thermo-chemistry of the dust, molecular, atomic and ionized species, formation and evolution of individually resolved proto-stars and main-sequence stars with stellar feedback, and non-ideal magnetohydrodynamics with kinetic effects. In this paper, we present the first infrared spectra from FORGE'd in FIRE by studying the dust torus with post-processing Monte Carlo radiative transfer simulations. Our main findings are as follows:

\begin{enumerate}
    \item \textbf{The dust torus is composed of the outer accretion disk and dusty gas tidally torn from the surrounding ISM complex.} In FORGE'd in FIRE, this region corresponds to a magnetically-dominated quasi-steady state super-Eddington accreting disk, without any significant star formation or fragmentation. The vertical scale height of $H/R \sim 0.1-0.6$ is sustained by trans-Alfvénic and highly super-sonic turbulent and magnetic support.
    \item \textbf{Early (pre-outflow) stages of the dust torus appear as a buried, Compton-thick AGN, with optical/UV emission only escaping along a few ($\sim$ 5\%) sightlines.} The resulting spectra have a hot dust component that peaks around 3-4 $\mu\mathrm{m}$. Significant IR reprocessing suppresses silicate emission features due to dust self-absorption. The local dusty ISM reservoir reprocesses this emission further, causing the emission to appear colder and producing silicate absorption lines.
    \item \textbf{There is significant azimuthal IR anisotropy due to significant dust reprocessing in the cold dusty stream from the ISM reservoir that fuels accretion.} The NIR varies by nearly an order of magnitude and the MIR still varies by a factor of a few. This anisotropy yields preferentially lower inferred AGN luminosity measurements in orientations associated with the cold dusty stream.
    \item \textbf{Sublimation structure is anisotropic and stratified in dust grain size and composition.} This stratification is primarily due to the intrinsic \citet{Netzer_1987} accretion disk anisotropy, but also arises from dust clumps and streams obscuring regions shielding that leads to colder dust in those directions. Clump and stream shielding is particularly noticeable at large Eddington ratios $L_\mathrm{AGN}/L_\mathrm{edd}$. 
    \item \textbf{The stratification in sublimation structure leads to extinction curves that greatly vary as a function of orientation in the unscattered emission.} However, these extinction curves significantly grayed in the full emergent spectrum due to many repeated scattering events when exiting the optically thick system.
    \item \textbf{In later stages, bipolar outflows that burrow bicone cavities through the dusty polar region can reproduce a standard type 1-like spectrum while retaining many of the other early-stage properties.} After bicone evacuation, the system maintains weak silicate emission/absorption features and enhances IR anisotropy since some radiation that otherwise would heat the cold dusty stream escapes the system.
\end{enumerate}

These results suggest a dust torus formation picture in which the early phases are dominated by a magnetically supported, geometrically thick outer accretion disk fed by dusty inflows from the circumnuclear ISM, producing a heavily buried, Compton-thick AGN with strong IR anisotropy. If an AGN-driven outflow is strong enough to excavate bipolar cavities, it can reveal a standard type 1/2 system while retaining IR anisotropy that reflect the direction of dusty feeding.

The simulations here represent the first attempt to create observable predictions from a fully self-consistent, quasar formation simulation the cosmological environment to scales significantly below the dust sublimation region, but there are several key directions to expand upon. In principle, it is possible to expand this post-processing radiation analysis directly to galactic scales ($\gtrsim 100$ pc), explicitly including the stellar population and diffuse ISM dust contributions, in order to place the AGN emission in the context of the host galaxy spectra and determine direct system observables. The broad line region is another largely unconstrained region that lies on the boundary between physics conveniently probed self-consistently by FORGE'd in FIRE, and would be a particularly interesting target for a radiative transfer analysis as there are no obvious gas clouds present in this system. Moreover, explicitly following the dynamics of the outflow through de-refinement would give insight into the dynamics and observable of the dust torus as well as the surrounding AGN system. We hope to explore these directions in future work.

\begin{acknowledgments}

The authors thank Christian Knigge, Matthew Temple, Kyle Kremer, Sam Ponnada, and Yashvardhan Tomar for insightful discussions and comments. The numerical calculations in this paper were run on the Texas Advanced Computing Center (TACC) allocation AST21010. J.B. acknowledges support from a Natural Sciences and Engineering Research Council of Canada (NSERC) doctoral scholarship. P.F.H. acknowledges support from a Simons Investigator Grant. 

\end{acknowledgments}

\software{
\href{https://github.com/SKIRT/SKIRT9}{\texttt{SKIRT}}: \cite{SKIRT_2020};
\href{https://github.com/yt-project/yt}{\texttt{yt}}: \cite{yt};
\href{https://www.numpy.org/}{\texttt{numpy}}: \cite{numpy}
\href{https://www.matplotlib.org/}{\texttt{matplotlib}}: \cite{matplotlib}
\href{https://www.scipy.org/}{\texttt{scipy}}: \cite{scipy}
\href{https://www.astropy.org/}{\texttt{astropy}}: \cite{astropy18}
\href{https://www.pandas.pydata.org/}{\texttt{pandas}}: \cite{pandas}
\href{https://scikit-image.org}{\texttt{scikit-image}}: \cite{skimage};
}

\bibliography{paper}{}
\bibliographystyle{aasjournal}

\appendix
\restartappendixnumbering

\section{Accretion Disk Spectrum}\label{app:input_spectrum}

In this section, we motivate the use of our accretion disk power-law component input spectrum that models the accretion disk. Considering that the exact shape of the spectrum has little impact on infrared emission, it is commonplace in dust reprocessing simulations to approximate the accretion disk SED as a series of power-law components \citep[e.g.][]{Granato_1994, Schartmann_2005, Honig_2010, Stalevski_2012}. Some simply model off of a \citet{Shakura_1973}-like thin disk emission spectrum, but more observationally constrained models consist of power-law components fitted to UV-optical mean composite quasar spectra \citep[e.g.][]{Zheng_1997, Vanden_Berk_2001, Scott_2004, Krawczyk_2013} cutoff by a Rayleigh-Jeans tail $\alpha_\nu = 2$ (here defined as $f_\nu \propto \nu^{\alpha_\nu}$) at some effective minimum accretion disk temperature, possibly with some extreme-UV and x-ray power-law component. In the bolometric quasar luminosity function (QLF) estimation literature \citep[e.g.][]{Richards_2006, Hopkins_2007, Shen_2020}, where more precise SED shape estimations are needed, it is common to use a mean quasar SED including dust torus emission along with an X-ray SED template with a cutoff power-law model using some photon index $\Gamma = 1 -\alpha_\mathrm{x}$ and a luminosity-dependent optical/UV to x-ray power-law component $\alpha_\mathrm{ox}$. While the x-ray component itself should not contribute much to the dust re-emission luminosity, the overall shape varies significantly enough between models that they can affect the bolometric luminosity normalization by a factor of a few. Since this could potentially affect the interpretation of some of our results, such as the sublimation radius, we construct an updated power-law input model for our post-processing simulation.

Our input accretion disk model consists of five components, modeled roughly on the spectrum used in \citep{Shen_2020} for QLF estimation. We adopt the commonly used $\alpha_\mathrm{o} = -0.5$ optical/UV thermal component from $0.1 \mu$m to $1\mu$m, consistent with \citet{Vanden_Berk_2001}, \citet{Richards_2006} and \citet{Krawczyk_2013}. We cut this off with a Rayleigh-Jeans tail with $T \sim 10^4$ K, corresponding to the inner limit of the accretion disk radiation temperature in FORGE'd in FIRE. On the other end of the thermal emission, near the Lyman limit, we use a $\alpha_\mathrm{u} = -1.75$ extreme UV component out to 500$~\text{\AA}$, consistent with \citet{Telfer_2002} and \citet{Lusso_2015}. We also have an X-ray component with $\Gamma = 1.9$ consistent with \citet{Dadina_2008} and \citet{Ueda_2014}, between 0.25 keV and $\sim$10 keV. We do not add any X-ray reflection component, since it only becomes significant at very hard X-rays, beyond the wavelengths we study here. We then scale this X-ray component such that the equivalent power-law between between $L_\nu$(2500$~\text{\AA}$) and $L_\nu$(2 keV) is $\alpha_\mathrm{ox} \sim -1.5$, consistent with \citet{Steffen_2006} for $\nu L_\nu \sim 10^{45.5}~\mathrm{erg}~\mathrm{s}^{-1}$ for 2500$~\text{\AA}$, corresponding to $L_\mathrm{bol} \sim 5\times 10^{45}$ in our model. At this luminosity, this scaling leads to a power-law component of $\alpha_\mathrm{ux} = -2.4$ between 50$~\text{\AA}$ and 500$~\text{\AA}$. Since the exact value of $\alpha_\mathrm{ox}$ is known to correlate non-linearly with luminosity \citep{Steffen_2006, Just_2007, Lusso_2010}, our power-law component between $50~\text{\AA} \leq \lambda \leq 500~\text{\AA}$ could in theory change with luminosity as $\alpha_\mathrm{ux}(L_\mathrm{AGN}) \sim 2.61\alpha_\mathrm{ox}(L_\mathrm{AGN}) + 1.52$ for some choice of $\alpha_\mathrm{ox}(L_\mathrm{AGN})$. However, on the luminosity scales that we study here, it only scales the X-ray luminosity on the order of $10^{-1}$ dex. This does not significantly impact total dust absorption or bolometric luminosity and thus would not significantly affect this study.

\begin{figure*}[t]
    \centering
    \includegraphics[width=\textwidth]{ 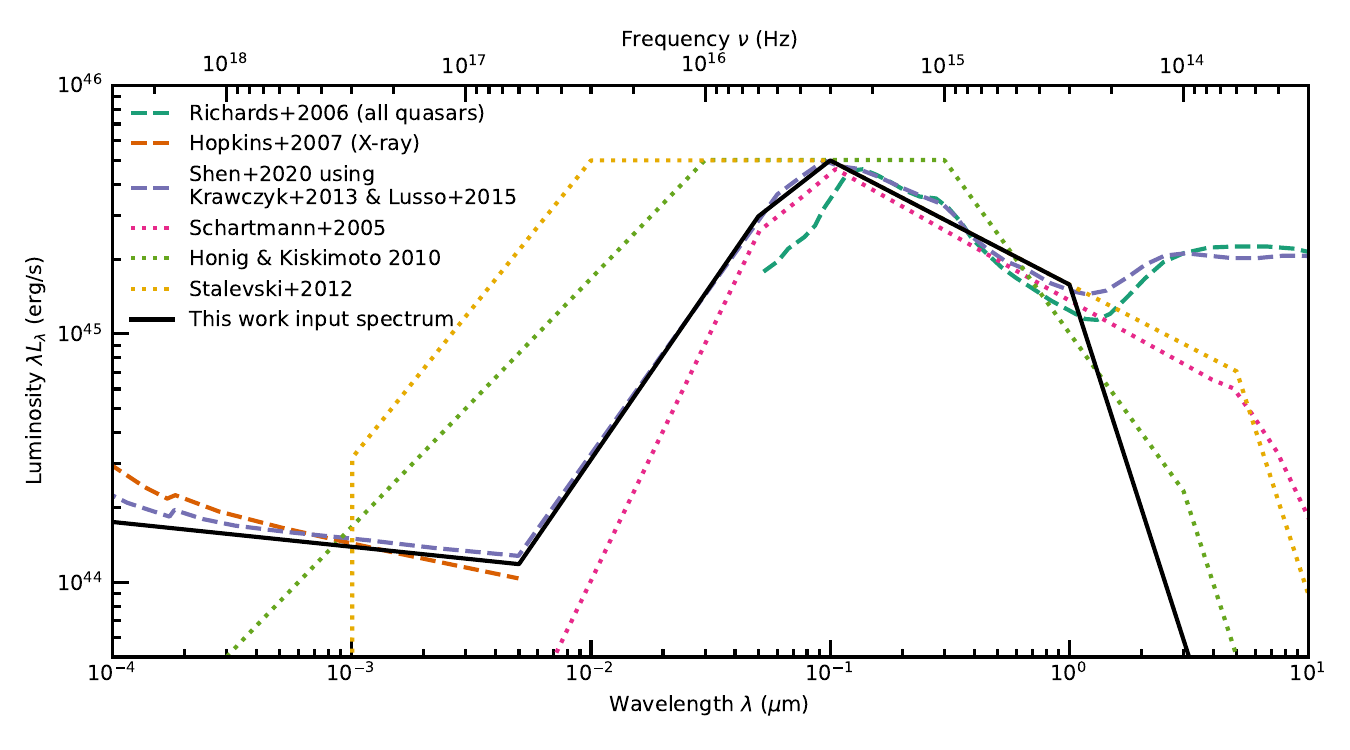}
    \caption{Our power-law component model input accretion disk spectrum (black solid line) compared to previous work. In the dashed lines, we show the \citet{Richards_2006}, \citet{Hopkins_2007} and \citet{Shen_2020} works that precisely estimate the mean quasar spectra for quasar luminosity function (QLF) estimation, but also include the infrared dust torus emission ($\lambda \gtrsim 1~\mu\mathrm{m}$). In the dotted lines, we show the input SED models used in radiative transfer simulations of AGN dust absorption and re-emission by \citet{Schartmann_2005}, \citet{Honig_2010} and \citet{Stalevski_2012}. Our input spectrum fits well with the non-dust torus components of the observationally constrained mean quasar spectra models used in QLF studies, and generally provides more accuracy in the X-ray and extreme UV regions than commonly used dust re-processing input spectra. The majority of the luminosity is in the optical/UV part of the spectrum, however, where all models roughly agree. }
    \label{fig:input_spectrum}
\end{figure*}

The full shape of our central AGN disk source SED is thus described by 
\begin{equation}
    \lambda L_{\lambda} \propto 
    \begin{cases} 
     \lambda^{-1/10} & 1~\text{\AA} \leq \lambda \leq 50~\text{\AA} \\  
      \lambda^{7/5} & 50~\text{\AA} \leq \lambda \leq 500~\text{\AA} \\  
      \lambda^{3/4} & 500~\text{\AA} \leq \lambda \leq 1000~\text{\AA} \\
      \lambda^{-1/2} & 1000~\text{\AA} \leq \lambda \leq 1~\mu\mathrm{m} \\
      \lambda^{-3} & \lambda \geq 1~\mu\mathrm{m}
   \end{cases}
\end{equation}
and is plotted in Figure \ref{fig:input_spectrum}. We show that our input spectrum is generally more consistent in the X-ray/extreme UV with the precise quasar mean spectra compiled for QLF studies (dashed) with $\alpha_\mathrm{ox} \sim -1.5$ than the traditional dust torus input spectra (dotted). However, we stress that there is very little discrepancy between any of the models in the optical/UV, where most of the dust-absorbed luminosity lies.

\end{document}